\documentclass{article}
\usepackage{amssymb}
\usepackage{hyperref}
\usepackage{caption}
\usepackage{graphicx}
\usepackage{epstopdf}

\usepackage[noend]{algpseudocode}
\usepackage{algorithm}
\usepackage{threeparttable}
\usepackage[table]{xcolor}
\newtheorem{df}{Definition}
\newtheorem{lem}[df]{Lemma}

\title{Hamiltonian cycles in hypercubes with faulty edges}
\author{Janusz Dybizba\'nski, Andrzej Szepietowski \\
\small Institute of Informatics,\\ 
\small Faculty of Mathematics, Physics and Informatics, \\
\small University of Gda\'nsk, 
\small 80-308 Gda\'nsk, Poland \\
\small \texttt{jdybiz@inf.ug.edu.pl, matszp@inf.ug.edu.pl}}

\date{}

\begin{document}

\maketitle

\begin{abstract}
Szepietowski [A. Szepietowski, {\sl Hamiltonian cycles in hypercubes with $2n-4$ faulty edges,} Information Sciences, 215 (2012) 75--82]
observed that the hypercube $Q_n$ is not Hamiltonian if it contains 
a trap disconnected halfway.
A proper subgraph $T$ is disconnected halfway if at least half of its nodes have parity 0
(or 1, resp.)
 and  the edges joining all nodes of parity 0 (or 1, resp.)  in $T$  with nodes outside $T$, are faulty.
 The simplest  examples of such traps are: 
(1) a vertex with $n-1$ incident faulty edges, or (2) a cycle $(u,v,w,x)$, where all edges going out of the
cycle from $u$ and $w$ are faulty.
In this paper we describe all traps disconnected halfway $T$ with the size $|T|\le8$,
and discuss the problem whether there exist small sets of faulty edges
which preclude Hamiltonian cycles and are not based on sets disconnected halfway. We describe heuristic which detects sets of faulty edges which preclude HC
also those sets that are not based on subgraphs disconnected halfway.
We describe all $Q_4$ cubes that are not Hamiltonian, and all $Q_5$ 
cubes with 8 or 9 faulty edges that are not Hamiltonian.
\end{abstract}


\section{Introduction}\label{intro}
\begin{figure}[htb]
\centering
\includegraphics{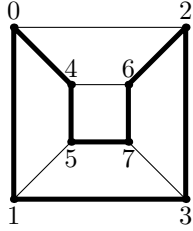}
\caption{The hypercube $Q_3$ with a Hamiltonian cycle.}
\label{figC4}
\end{figure}

An $n$-dimensional hypercube ($n$-cube), denoted by $Q_n$, is an undirected
graph with $2^n$ vertices each labeled with a distinct binary string of length $n$.
We shall identify the vertices of $Q_n$ with integers in $\{0,\dots, 2^n-1\}$ in the usual way, see Fig.~\ref{figC4}.
Two vertices are connected by an edge if and only if their labels differ in one
bit. The hypercube is one of the most popular interconnecting networks.
It has several excellent properties such as: recursive structure, symmetry,
small diameter, low degree and easy routing algorithm, which are important
when designing parallel or distributed systems.
 The hypercube $Q_n$ is  bipartite, the set of nodes is the union
of two sets: nodes of parity 0 (the number of ones
in their labels is even), and nodes of parity 1 (the number of ones
is odd), and each edge connects nodes of different parity.
The hypercube is Hamiltonian, i.e.  it contains a cycle
which visits each node in the cube exactly once, see Fig.~\ref{figC4}.
An important property of the hypercube is its fault tolerance.
It contains Hamiltonian cycles even if some edges or vertices are faulty.

In this paper we consider Hamiltonian cycles in hypercube $Q_n$ with faulty edges
(we shall denote the set of faulty edges  by $F$). 
For $n\le 2$, the problem is trivial, so in the sequel we assume that $n\ge3$.
It is obvious that if the cube with the set of faulty edges $F$ is not
Hamiltonian, then also the cube with the set of faulty edges $G$ is not Hamiltonian,
if $F\subset G$. We say that the set $F$ is minimal if no proper subset of $F$ precludes  Hamiltonian cycles; and the cube with a set of faulty edges $F$ is minimal if $F$ is minimal. 

Tsai et al.~\cite{3}
showed that if there are no more than $n-2$ faulty edges in
the hypercube $Q_n$, then it contains a Hamiltonian cycle.
The hypercube $Q_n$ is $n$-regular, so it is not Hamiltonian if it contains
a vertex with incident $n-1$ faulty edges. We shall call such a set of faulty edges a local trap.
In  \cite{4} Xu et al. proposed, so called, conditional fault model, where each vertex is incident with at least two healthy edges.
Under this model, they showed that with $n-1$ faulty edges the cube is
still Hamiltonian. Chan and Lee~\cite{1} showed  that also with
$|F|\le 2n-5$  faulty edges $Q_n$ is Hamiltonian.
Shih et al.~\cite{2}  observed that the bound $2n-5$ is optimal, because with
$2n-4$ faulty edges it is possible to build another kind
of local trap, namely a (healthy) cycle $(u,v,w,x)$, where all edges going out of the
cycle from $u$ and $w$ are faulty. This kind of trap is called $f_4$-cycle in~\cite{Li}.
Szepietowski~\cite{Sz} observed that these two kinds of traps are
examples of a more general scheme, namely subgraphs disconnected halfway.
\begin{df}\label{defdhw}
A proper subgraph $T\subset Q_n$ is disconnected halfway if either
(1) at least half of the nodes of $T$ have parity 0
 and  the edges joining all nodes of parity 0 in $T$  with nodes outside $T$, are faulty,
 or
 (2) at least half of the nodes of $T$ have parity 1
 and  the edges joining all nodes of parity 1 in $T$  with nodes outside $T$, are faulty.
\end{df}

There is a different definition of  a set disconnected halfway in \cite{Sz}, where
exactly half of the nodes have parity 0.  In Section~\ref{dhw} we shall show that
if the trap is minimal then  half of the nodes in $T$  have parity 0. Since
we are mainly interested in minimal traps, these definitions are equivalent.
We  identify a set $T$  of vertices with the subgraph induced by the set.
We shall say that two sets $T_1$ and $T_2$ are isomorphic if there is an
automorphism of $Q_n$ which maps $T_1$ onto $T_2$. Similarly, we shall say that two sets of faulty edges $F_1$ and $F_2$ are isomorphic if there is an
automorphism of $Q_n$ which maps $F_1$ onto $F_2$.

\begin{lem}\label{lem2}[see \cite{Sz}] The cube 
$Q_n$ is not Hamiltonian if it has a trap disconnected halfway.
\end{lem}
\noindent{\bf Proof.}
Any cycle which does not visit any node twice, when visiting $T$
enters and leaves it in  nodes of parity 1, so during one visit it goes
through $i$ nodes of parity 0 and $i+1$ nodes of parity 1, for some $i$.
Hence, the cycle cannot visit all nodes.\hfill $\Box$

Observe that one node with $n-1$ faulty incident edges forms
with its neighbor a subcube of dimension 1 disconnected halfway. We shall call it the cube $Q_1$ disconnected halfway, and denote by $Q_1$-DHW. If $T$ is the
cycle $C_4=(u,v,w,x)$ (subcube of dimension 2), then we obtain the trap described in \cite{2} and mentioned above. 
We shall call it the cube $Q_2$ disconnected halfway, and denote by $Q_2$-DHW. Similarly, with $4(n-3)$ faulty edges we can disconnect halfway the 3-dimensional subcube $Q_3$
and with $8(n-4)$ faulty edges the 4-dimensional subcube $Q_4$.
We shall denote these traps by $Q_3$-DHW and $Q_4$-DHW respectively. Table~\ref{tab2} contains the numbers of faulty edges $|F|$ needed to
build in $Q_n$ traps based on sets disconnected halfway $T$,
for $|T|\le 6$, and $3\le n\le 7$.

Szepietowski~\cite{Sz} proved that  the cube $Q_n$ with  $n\ge5$ and with $2n-4$ faulty edges   is not Hamiltonian only if it  contains a $Q_1$-DHW or $Q_2$-DHW trap.
Liu and Wang \cite{Liu} showed that the same is true for hypercubes with $3n-8$ 
faulty edges.
The bound $3n-8$ is sharp, because with $3n-7$   faulty edges it is possible to build another kind of local trap, namely a (healthy) cycle $(u,v,w,x,y,z)$, where all edges going out of the cycle from $u$, $w$, and $y$ are faulty, see \cite{Li,Liu,Sz}. 
We shall call this trap a $C_6$-DHW trap.
Recently Li et al.~\cite{Li} proved that with $n\ge 6$ and with $3n-7$ faulty edges
$Q_n$ is not Hamiltonian only if it  contains
a $Q_1$-DHW, $Q_2$-DHW, or $C_6$-DHW trap.

\begin{figure}[htb]
\centering
\includegraphics{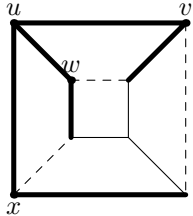}
\caption{The CL trap in $Q_3$.}
\label{figCLtrap}
\end{figure}

In \cite{Sz}, another kind of trap was described which is not a set disconnected halfway. Consider the cube $Q_3$ with three faulty edges as shown in 
Fig.~\ref{figCLtrap}.
The node $u$ has three neighbors $v$, $w$ and $x$, each of degree 2, i.e. with only
two healthy incident edges. The cube is not Hamiltonian.
Otherwise the Hamiltonian cycle going through $v$, $w$ and $x$ must use
edges $(u,v)$, $(u,w)$ and $(u,x)$ which is impossible.
Similar traps we can build for $n\ge5$. We only need three healthy edges
$(u,v)$, $(u,w)$ and $(u,x)$, where $v$, $w$,  and $x$ are of degree 2 in $Q_n-F$.
In this way we  obtain  traps with
$3(n-2)=3n-6$ faulty edges. We shall call them  CL (claw) traps.
Table~\ref{tab1} presents the upper bounds mentioned above, for n-cubes with $3\le n\le7$,
and information about Hamiltonian cycles.

\begin{table}
\centering
\begin{threeparttable}
\begin{tabular}{c||p{1cm}|p{1cm}|p{1cm}|p{1cm}|p{2cm}||c}
$|F|$ & \multicolumn{5}{c||}{dimension} & ref\\
 & $3$ & $4$ & $5$ & $6$ & $7$ & \\
\hline
\hline
$\leq n-2$ & $1$ & $2$ & $3$ & $4$ & $5$ & \cite{3}\\
& \multicolumn{5}{p{8.71cm}||}{\cellcolor{gray!25} \centering cube is Hamiltonian} & \\
\hline
$\leq 2n-5$ & $1$ & $3$ & $5$ & $7$ & $9$ & \cite{1}\\
& \tnote{*} & \multicolumn{4}{p{7.27cm}||}{\cellcolor{gray!25} \centering cube is Hamiltonian provided it contains no $Q_1$-DHW} & \\
\hline
$\leq 3n-8$ & $1$ & $4$ & $7$ & $10$ & $13$ & \cite{Liu}\\
& \tnote{*} & \tnote{**} & \multicolumn{3}{p{5.84cm}||}{\cellcolor{gray!25} \centering cube is Hamiltonian  provided it contains neither $Q_1-$DHW, nor $Q_2$-DHW} & \\
\hline
$= 3n-7$ & $2$ & $5$ & $8$ & $11$ & $14$ & \cite{Li}\\
 & \tnote{***} & \tnote{****} & \tnote{*****}  & \multicolumn{2}{p{4.4cm}||}{\cellcolor{gray!25} \centering cube is Hamiltonian provided it contains neither $Q_1-$DHW, nor $Q_2$-DHW, nor $C_6$-DHW} & \\
\end{tabular}
\begin{tablenotes}
\item[*] $Q_3$ with $|F|\leq 1$ faulty edges is Hamiltonian
\item[**] $Q_4$ with $|F|\leq 4$ is Hamiltonian  provided it contains neither $Q_1$-DHW, nor $Q_2$-DHW, nor $Q_3$-DHW 
\item[***] $Q_3$ with $|F|\leq 2$ is Hamiltonian provided it contains no $Q_1$-DHW
\item[****] $Q_4$ with $|F|\leq 5$ is Hamiltonian provided it contains neither $Q_1$-DHW, nor $Q_2$-DHW, nor $Q_3$-DHW, nor $C_6^1$-DHW. See Section~\ref{Q4}.
\item[*****] $Q_5$ with $|F|\leq 8$ is Hamiltonian provided it contains neither $Q_1$-DHW, nor $Q_2$-DHW, nor $Q_3$-DHW, nor $Q_4$-DHW, nor $C_6^1$-DHW. See Section~\ref{Q5}.
\end{tablenotes}
\caption{The upper bounds described in~\cite{3,1,Liu,Li}, on the number of faulty edges, for $3\leq n \leq 7$}
\label{tab1}
\end{threeparttable}
\end{table}

Each of the trap mentioned above consists of small collection of faulty edges which is easy to detect. In this paper we consider other traps of this kind. In Sections~\ref{dhw}--\ref{C8}
we concentrate on small sets disconnected halfway. In Section~\ref{dhw} we show that if 
$|T|\le 8$ then we can consider only subcubes $Q_1$, $Q_2$, $Q_3$ and cycles of length 6 and 8, because if $T$ is minimal and $|T|\le8$ then it is $Q_1$ or contains a cycle $C_4$, $C_6$, $C_8$, or contains a CL-trap.

In Section~\ref{SecC6} we consider traps disconnected halfway where 
the subgraph $T$ is a cycle of length six.
We show that two nonisomorphic traps can be build in this way.
We shall denote them by $C_6^1$-DHW and $C_6^2$-DHW. As we have already mentioned,
$C_6^1$-DHW is described in \cite{Li,Liu,Sz}.
In Section~\ref{C8} we consider traps disconnected halfway where 
the subgraph $T$ is a cycle of length eight. There are seven nonisomorphic
cycles of length 8 and nine nonisomorphic traps can be build by cutting off
their vertices. We shall denote them by $C_8$-DHW traps.
Tables~\ref{tab2} and ~\ref{tab3} present all traps disconnected halfway with $|T|\le8$

\begin{table}
\centering
\begin{threeparttable}
\begin{tabular}{c||c||c|c|c|c|c||c}
trap & $|F|$ & \multicolumn{5}{|c||}{dimension} & figure\\
 & & $3$ & $4$ & $5$ & $6$ & $7$ & \\
\hline
\hline
$Q_1$-DHW & $n-1$ & $2$ & $3$ & $4$ & $5$ & $6$ & \\
$Q_2$-DHW & $2n-4$ & $2$ & $4$ & $6$ & $8$ & $10$ & \\
$Q_3$-DHW & $4n-12$ & --- & $4$ & $8$ & $12$ & $16$ & \\
$Q_4$-DHW & $8n-32$ & --- & --- & $8$ & $16$ & $24$ & \\
\hline
$C_6^1$-DHW & $3n-7$ & {\tiny not minimal} & $5$                 & $8$ & $11$ & $14$ & Fig.~\ref{figC6c1trap}\\
$C_6^2$-DHW & $3n-6$ & {\tiny not minimal} & {\tiny not minimal} & $9$ & $12$ & $15$ & Fig.~\ref{figC6c2trap}\\
$CL$ & $3n-6$    & 3                   & $6$                 & $9$ & $12$ & $15$ & Fig.~\ref{figCLtrap}\\
\end{tabular}
\caption{The numbers of faulty edges $|F|$ needed to build in $Q_n$ traps based on sets disconnected halfway $T$, for $|T|\le 6$, and $3\le n\le 7$. In all these cases, except CL trap, two traps: one obtained by cuting off vertices with parity 0, and the other obtained by cuting off vertices with parity 1 are isomorphic.
}
\label{tab2}
\end{threeparttable}
\end{table}

\begin{table}
\centering
\begin{threeparttable}
\begin{tabular}{p{2cm}||c||c|c|c|c||c}
trap & $|F|$ & \multicolumn{4}{|c||}{dimension} & figure\\
 & & $4$ & $5$ & $6$ & $7$ & \\
\hline
\hline
$C_8^1$-DHW \tnote{****} & $4n-12$ & $4$ & $8$ & $12$ & $16$ & \\
$C_8^2$-DHW \tnote{***} & $4n-9$ & {\tiny not minimal} & $11$ & $15$ & $19$ & Fig.~\ref{figC8c2trap}\\
$C_8^3$-DHW \tnote{***} & $4n-10$ & $6$ & $10$ & $14$ & $18$ & Fig.~\ref{figC8c3trap} \\
$C_8^4$-DHW \tnote{***} & $4n-10$ & $6$ & $10$ & $14$ & $18$ & Fig.~\ref{figC8c4trap}\\
$C_8^5$-DHW \tnote{***} & $4n-8$ & $8$ &                $12$ & $16$ & $20$ & Fig.~\ref{figC8c5trap}\\
$C_8^6$-DHW \tnote{*} & $4n-8$ & {\tiny not minimal} & $12$ & $16$ & $20$ & Fig.~\ref{figC8c6trap}\\
$C_8^6$-DHW \tnote{**} & $4n-8$ & {\tiny not minimal} & $12$ & $16$ & $20$ & \\
$C_8^7$-DHW \tnote{*} & $4n-8$ & {\tiny not minimal} & {\tiny not minimal} & $16$ & $20$ & Fig.~\ref{figC8c7trap}\\
$C_8^7$-DHW \tnote{**} & $4n-8$ & {\tiny not minimal} & $12$ & $16$ & $20$ & \\
\end{tabular}
\begin{tablenotes}
\item[*] we cut off vertices with parity 0,
\item[**] we cut off vertices with parity 1,
\item[***] Two traps: one obtained by cutting off vertices with parity 0, and the other obtained by cutting off vertices with parity 1 are isomorphic,
\item[****] The cycle $C_8^1$ is the subcube $Q_3$. Hence $C_8^1$-DHW is $Q_3$-DHW.
\end{tablenotes}
\caption{The numbers of faulty edges $|F|$ needed to build in $Q_n$ traps based on cycles of length 8 disconnected halfway, for $4\le n\le 7$.}
\label{tab3}
\end{threeparttable}
\end{table}

In Section~\ref{H} we discuss the problem whether there exist small sets of faulty edges
which preclude Hamiltonian cycles and are not based on sets disconnected halfway. First examples of such traps are CL traps, where we have three vertices of degree 2 with a common neighbour. In Section~\ref{H} we describe heuristic which detects sets of faulty edges which preclude Hamiltonian cycle. The heuristic is not perfect. There are $Q_4$ cubes with 8 or more faulty edges that are not Hamiltonian, do not contain any traps of the kinds described above,
and cannot be detected by the heuristic.

In the second part of the paper we concentrate on nonhamiltonian cubes $Q_n$
with $n\le5$.
In $Q_3$ there is, up to isomorphism, only one Hamiltonian cycle, see Fig.~\ref{figC4}. It is easy to check that $Q_3$ is not Hamiltonian only
if it contains a $Q_1$-DHW, $Q_2$-DHW, or CL trap, see \cite{Sz}.

In Section~\ref{Q4} we describe all nonhamiltonian $Q_4$ cubes.
Nonhamiltonian $Q_4$ cubes with no more than four faulty edges are described in 
 ~\cite{3,4,Sz}. Using computer we found that: the cube  $Q_4$ with  5 faulty edges is not Hamiltonian
only if it contains a $Q_1$-DHW, $Q_2$-DHW, $Q_3$-DHW, or $C_6^1$-DHW trap, and
with  6 faulty edges $Q_4$ is not Hamiltonian
only if it contains a $CL$, $Q_1$-DHW, $Q_2$-DHW, $Q_3$-DHW, $C_6^1$-DHW, $C_8^3$-DHW, or $C_8^4$-DHW trap.
There are 25 nonisomorphic $Q_4$ cubes with 7  faulty edges that are not Hamiltonian and do not contain any traps of the kinds described above.
All these cubes can be recognized by the heuristic described in Section~\ref{H}.
There are $Q_4$ cubes with 8 or more faulty edges that are neither Hamiltonian, nor contain any traps of the kinds described above,
nor can be detected by the heuristic. We present them in Appendix.

In Section~\ref{Q5} we describe nonhamiltonian $Q_5$ cubes.
Nonhamiltonian $Q_5$ cubes with no more than seven faulty edges are described in 
 ~\cite{3,4,Sz,Liu}. Using computer we found that:

\begin{itemize}
\item with  8 faulty edges $Q_5$ is not Hamiltonian
only if it contains a $Q_1$-DHW, $Q_2$-DHW, $Q_3$-DHW, $Q_4$-DHW, or $C_6$-DHW trap,
\item with  9 faulty edges $Q_4$ is not Hamiltonian
only if it contains a $Q_1$-DHW, $Q_2$-DHW, $Q_3$-DHW, $Q_4$-DHW, $C_6^1$-DHW, $C_6^2$-DHW, or $CL$ trap.
\end{itemize}
We are not able to give full description of the $Q_5$ cubes with ten faulty edges that
are not Hamiltonian. With ten faulty edges we can build a trap disconnected halfway with a cycle of length 8.
At the end of Section~\ref{Q5} we show that with ten faulty edges we can build in $Q_5$ three other types of traps
disconnected halfway.

\section{Preliminaries}\label{pre}
An $n$-dimensional hypercube (cube), denoted by $Q_n$, is an undirected
graph with $2^n$ nodes, each labeled with a distinct binary string
$b_{n-1}\dots b_1b_0$, where $b_i\in\{0,1\}$. A vertex $x=b_{n-1}\dots b_i\dots b_0$
and the vertex $x^{(i)}=b_{n-1}\dots \bar{b_i}\dots b_0$ are connected by an edge along
dimension $i$, where $\bar{b_i}$ is the negation of $b_i$.
 The hypercube $Q_n$ is  bipartite, the set of nodes is the union
of two sets: nodes of parity 0 (the number of ones
in their labels is even), and nodes of parity 1 (the number of ones
is odd), and each edge connects nodes of different parity.

There are two kinds of symmetries in $Q_n$. Firstly, for each
vertex $v=v_{n-1}\dots v_0$, there is an automorphism $h_v$ of $Q_n$ 
defined by $h_v(x)= x\, \oplus \, v$ more precisely, $h_v(x_{n-1}\dots x_0)=w_{n-1}\dots w_0$, where
$w_i=x_i+v_i\bmod 2$, for every $i$ satisfying: $0\le i\le n-1$.
The second kind of automorphism permutes dimensions.
More precisely,
for every permutation $\pi$ acting on $\{0,\dots,n-1\}$, there is an automorphism $g_{\pi}$ of $Q_n$ defined by the rule
$g_{\pi}(x_{n-1}\dots x_0)=x_{\pi(n-1)}\dots x_{\pi(0)}$.
Each automorphism of $Q_n$ is a composition of two automorphisms of the kinds described above~\cite{Chen}.

For $i\in\{0,\dots ,n-1\}$, the $i$-partition of $Q_n$ (or partition along
the $i$-th dimension) is the partition into two  disjoint $(n-1)$-subcubes
$Q^L_{n-1}=\{x\in Q_n\mid x_i=0\}$ and
$Q^R_{n-1}=\{x\in Q_n\mid x_i=1\}$.
We shall also write that the cube $Q_n$ is decomposed into two
subcubes $Q^L_{n-1}$ and $Q^R_{n-1}$ along the $i$-th dimension,
or simply into $Q^L$ and $Q^R$.
Let $D_i$ denote the set of edges of dimension $i$. These are edges between $Q^L$ and $Q^R$ and they are called crossing edges.
We shall say that a crossing edge $(u,u^{(i)})$, with $u\in Q^L$ is even  if the vertex $u$ is of parity 0, and is odd, if $u$ is of parity 1.

\begin{lem}\label{cycle}
Consider a cycle $C= (v_0,\dots, v_{2k-1})$ of length $2k$ in $Q_n$.
Let $(e_1, \dots, e_{2k})$ be the sequence of edges in $C$,
where $e_j=(v_{j-1},v_j)$, for every $1\le j \le 2k-1$,
and $e_{2k}=(v_{2k-1}, v_0)$. Let $(d_1, \dots, d_{2k})$ be the sequence of dimensions of the edges in $C$, with $d_j$ being the
dimension of $e_j$. 
Then:

(1) No two incident edges go in the same dimension.

(2) Each dimension $i\in \{0,\dots ,n-1\}$ appears in $C$ an even number of times.
Hence, there are at most $k$ different dimensions in $C$.

(3) The cycle $C$ is a subset of a subcube of dimension $k$. More precisely, there is an automorphism of $Q_n$ mapping the 
cycle into the subcube $\{x\mid 0\le x<2^k\}$.

(5) If $C$ has no edges in dimension $i$, then $C$  is contained either in $Q^L_{n-1}$ or in $Q^R_{n-1}$,
where $Q^L_{n-1}$ and $Q^R_{n-1}$ are formed after partition of $Q_n$  along the $i$-th dimension.
\end{lem}

Note that two cycles with the 
same sequence of dimensions are isomorphic, i.e. there is an automorphism of $Q_n$ that maps the first cycle onto the other.
\begin{lem}\label{cycle2}
Suppose that we have a Hamiltonian cycle $C$ and partition of $Q_n$ along dimension $i$, and let $e_1$, $e_2$, ... ,$e_s$
are all crossing edges that belong to $C$.

Then: $s$ is even, $s>0$, and the set of edges is balanced, i.e. half of these edges are even and the other half are odd.
\end{lem}

\noindent{\bf Proof:} From Lemma~\ref{cycle} it follows that $s$ is even and positive.
Suppose that, for each $i$, $1\le i\le s$, we have 
$e_i=(u_i,v_i)$ with $u_i\in Q^L$. Vertices
$u_1$, $u_2$, ... ,$u_s$ are connected in $Q^L$ by fragments of $C$. 
We may assume that, there are paths $P_1$, $P_3$, ... ,$P_{s\over2}$,
such that $P_i$ connects $u_{2i-1}$ with $u_{2i}$, for each $i$, such that $1\le i \le {s\over2}$.
These paths are disjoint and they cover the whole $Q^L$.
Hence, by \cite{caha}, their ends are balanced, half of them have parity one and the other half parity zero.
\hfill $\Box$

\section{Traps disconnected halfway}\label{dhw}
Suppose that $T$ is a proper subset of the set of vertices in $Q_n$. We identify the set with the subgraph induced  by $T$. 
Let $|T|_0$ denote the number of nodes with parity 0 in $T$
and $|T|_1$ denote the number of nodes with parity 1 in $T$.
The set $T$ is balanced if  $|T|_0=|T|_1$ .
By Definition~\ref{defdhw},  the set $T$ is disconnected halfway if either
(1) $|T|_0\ge|T|_1$
 and  the edges joining all nodes of parity 0 in $T$  with nodes outside $T$, are faulty,
 or
 (2) $|T|_0\le|T|_1$
 and  the edges joining all nodes of parity 1 in $T$  with nodes outside $T$, are faulty.
By Lemma~\ref{lem2},  the cube $Q_n$,  is not Hamiltonian if it contains
a trap $T$ disconnected halfway.
In this section we only consider  the first kind of traps. By symmetry, all 
considerations will apply to the second kind of traps.

Let $F$ be the set of faulty edges in $Q_n$
and $T$ be a set disconnected halfway,  with $|T|_0 \ge|T|_1$.
If $|T|=1$, then the trap is not minimal. If $|T|=2$ and $T$ is connected and balanced, then we have $Q_1$-DHW, otherwise the trap is not minimal.
In the rest of this section we shall only consider traps with $|T|\ge3$.

\begin{lem}\label{L1}
Let $F$ be a set of faulty edges in $Q_n$
and $T$ be a set disconnected halfway,  where $|T|\ge3$, the edges joining all nodes of parity 0 in $T$  with nodes outside $T$, are faulty,
and $|T|_0 \ge|T|_1$.
\begin{enumerate} 
\item If there are two vertices  $u,v\in T$ such that  $(u,v)\in F$,
then $F$ is not minimal.
\item If $|T|_0 >|T|_1$, then $F$ is not minimal.
\item If $T$ is not connected  then $F$ is not minimal.
\item If $T$ has a pendant vertex $v$ (of degree 1), then $F$ is not minimal or there 
is a proper subset $T'\subset T$ which is disconnected halfway with the same 
set of faulty edges $F$.
\item Suppose that there is a partition of $T$ into two connected and balanced subsets $T_1$ and $T_2$
and two vertices $v_1\in T_1$ and $v_2\in T_2$, such that
$(v_1,v_2)$ is an edge, and every path in $T$ going from $v_1$ to $v_2$
contains the edge $(v_1,v_2)$. Then $F$ is not minimal.
In such a case we shall say that $(v_1,v_2)$ is a bridge between two balanced subsets.
\end{enumerate}
 \end{lem}
\noindent{\bf Proof.}

(1)
$T$ is DHW with the set of faulty edges $F-\{(u,v)\}$. In the rest of the proof
we assume that  there are no faulty edges between two nodes in $T$. 

(2) Let $x\in T$ be of parity 0 with incident faulty edges.
Remove from $F$ all edges incident to $x$. Now, $T-\{x\}$ is DHW.
In the rest of the proof we assume that $|T|_0=|T|_1$.

(3) One of the component is a trap DHW.
 
 (4) Let $v_0\in T$  be a neighbor of $v$.
 If $v$ is of parity 0, then $v$ and $v_0$ form a $Q_1$-DHW.
 If $v$ is of parity 1 ($v_0$ is of parity 0), then add the edge 
 $(v_0,v)$ to $F$. Now $T-\{v\}$ is a trap DHW.
 By the proof of (2), we can remove $v_0$ from $T$ and 
 all faulty edges incident to $v_0$ from $F$.  Note that at the end the edge
 $(v_0,v)$ is not in $F$ . 
 
 (5) We may assume that neither $v_1$ nor $v_2$ is pendant.
 One of the two vertices, say $v_2$, is of parity 1, and the edge $(v_1,v_2)$
 is the only edge between $T_1$ and $T_2$. These mean that  $T_2$
 is DHW.
 \hfill $\Box$

\bigbreak
In the sequel we will only consider minimal cubes.
Hence, we can assume that the set $T$ is connected, balanced (so with $|T|$ even),  without pendant vertices  and bridges
between balanced subsets.

\begin{lem}
Let $Q_n$ be a minimal cube with a set $T$ disconnected halfway, and no proper subset of $T$ is disconnected halfway with the same set of faulty edges.
Then
\begin{enumerate}
\item For $|T|=4$, we have a trap $Q_2$.-DHW.
\item For $|T|=6$, the set $T$ contains a cycle $C_6$.
\item For $|T|=8$, the set contains a cycle $C_8$, or $Q_n$ has a CL trap.
\end{enumerate}
\end{lem}
\noindent{\bf Proof.} $T$ does not have pendant vertices, hence it contains a cycle. If it contains a cycle $C_4$, say $(a,b,c,d)$, then by Lemma~\ref{cycle}, the cycle is a subcube $Q_2$. Thus, we have proved (1).

(2) For $|T|=6$, the set  $T=\{a,b,c,d,e,f\}$ contains a cycle $C_6$ or a subcube  $Q_2$, say $(a,b,c,d)$. In the later case both vertices $e$ and $f$ have only one neighbor in the subcube each. Since  $e$ and $f$ are not pendant,
they form the edge $(e,f)$ which is connected with an edge in the subcube.
More precisely, there is and edge $(u,v)$ in $(a,b,c,d)$ such that $(u,e)$ and $(v,f)$
are healthy edges.
Hence,  we have a cycle $C_6$.

(3)
For $|T|=8$, first we show that $T$ contains a cycle $C_8$ or  $C_6$.
The set $T$  contains a cycle $C_8$, or $C_6$, or a subcube  $Q_2$, say $(a,b,c,d)$. In the later case, since $T$ does not have a bridge between balanced subsets,  there is a path joining two vertices in
the subcube and going outside the subcube. It is not possible that the path has the length 2, so
it is longer than 2 and forms with vertices in $(a,b,c,d)$ a cycle of length 6 or 8. Suppose now that $(a,b,c,d,e,f)$ is a cycle in $T$. By Lemma~\ref{cycle},
the cycle is contained in a subcube $Q_3$, and in the next section
we show that there are only two kinds of $C_6$, see Fig.~\ref{figC6c1trap} and Fig~\ref{figC6c2trap}.
If the vertices $g$ and $h$ are also in the subcube, then we have 
$Q_3$-DHW. If we exclude pendant vertices, then $g$ and $h$ form
the edge $(g,h)$ and this edge is connected with one of the edges 
in $\{a,b,c,d,e,f\}$.
There are two cases:

Case 1. The edge $(g,h)$ is connected with an edge of the cycle,
and we have a cycle of length 8, see Fig.~\ref{figC8c2trap}, \ref{figC8c3trap}, \ref{figC8c4trap}.

Case 2. The edge $(g,h)$ is connected with a diagonal. This is possible if the cycle is $C_6^1$ depicted on Fig.~\ref{figC6c1trap} and the diagonal is $(0,2)$.
Now $T$ contains neither a pendant  vertex nor a bridge,
but if we cut off vertices of parity 0 (or 1) we obtain the CL-trap
in $Q_4$ and a trap which is not minimal if $n\ge5$ (it contains CL trap as a subgraph). \hfill $\Box$

\section{Cycles $C_6$ disconnected halfway}\label{SecC6}
\begin{figure}[htb]
\centering
\includegraphics{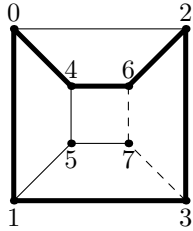}
\caption{The cycle $C_6^1$.}
\label{figC6c1trap}
\end{figure}

In this section we consider traps disconnected halfway where 
the subgraph $T$ is a cycle $C=(v_0,v_1, v_2,v_3,v_4,v_5)$ of length $6$ in $Q_n$, with $n\ge 3$.
Let $(d_1, d_2,d_3,d_4,d_5, d_6)$ be the sequence of dimensions of edges in $C$. By Lemma~\ref{cycle}, the cycle $C$ is contained in some 3-dimensional subcube, 
say $\{x\in Q_n\mid  x<8\}$. Moreover, $C$ is not contained in any 2- dimensional cube. Hence, for each $i\in\{0,1,2\}$, it has
 a pair of edges in $i$ dimension. We have two cases:

Case 1. There are two edges of the same dimension separated by one edge. 
Up to the isomorphism we have one possible sequence of dimensions $(0,1,0,2,1,2)$ and one cycle $C_6^1=(0,1,3,2,6,4)$, see Fig.~\ref{figC6c1trap}.
If we cut off vertices $0,3,6$ (or $1,2,4$)  from the vertices outside the cycle, then we obtain the $f_6$-cycle described in \cite{Li} or $C_6$-DHW trap from \cite{Sz}. 
We need $3n-7$ faulty edges to do this.
Note that the vertex $7$ is of degree one in $Q_3$. But if we embed
the trap in $Q_n$ with $n\ge 4$, then $7$ has degree at least two, see Table~\ref{tab2}.

\begin{figure}[htb]
\centering
\includegraphics{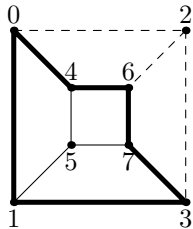}
\caption{The cycle $C_6^2$.}
\label{figC6c2trap}
\end{figure}

Case 2. Each pair of two edges of the same dimension is separated by two other edges, 
Up to the isomorphism, we have one possible sequence of dimensions $(0,1,2,0,1,2)$ and one cycle $C_6^2=(0,1,3,7,6,4)$, 
 see Fig.~\ref{figC6c2trap}.
In this case we need $3n-6$ faulty edges to cut off vertices $0,3,6$ from the vertices outside the cycle. 
Note that the vertex $2$ is of degree zero in $Q_3$. But if we embed
the trap in $Q_n$ with $n\ge 5$, then $2$ has degree at least two.
We get an isomorphic trap, if we cut off the vertices $1,7,4$, see Table~\ref{tab2}.

\section{Cycles $C_8$ disconnected halfway}\label{C8}
In this section we consider traps disconnected halfway where 
the subgraph $T$ is a cycle $C=(v_0,v_1, \dots, v_7)$ of length $8$ in $Q_n$. For $n=3$, $T$ is not a propers subset  of $Q_3$, so we assume that $n\ge 4$.
Let $(d_1, \dots, d_8)$ be the sequence of dimensions of edges in $C$. By Lemma~\ref{cycle}, the cycle $C$ is contained in some 4-dimensional subcube, say 
$\{x\in Q_n\mid x<16\}$. If $C$ is contained in some 
3-dimensional subcube $Q_3$, then it forms the Hamiltonian cycle in $Q_3$ (there is only one nonisomorphic  Hamiltonian cycle in $Q_3$). If we cut off every second vertex of the cycle from the vertices outside the cycle,
then we get a $Q_3$-DHW trap.
Assume now that $C$ is not contained in any 3-dimensional subcube.
Then, by Lemma~\ref{cycle}, it has a pair of edges in each of the four dimensions $\{0,1,2,3\}$.
We have three cases.

\begin{figure}[htb]
\centering
\includegraphics{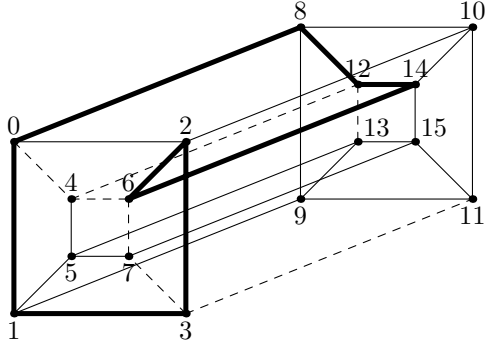}
\caption{The cycle $C_8^2$. The vertices of parity 0 are cut off. If we cut off the vertices with parity 1, then we obtain the isomorphic trap.}
\label{figC8c2trap}
\end{figure}

\begin{figure}[htb]
\centering
\includegraphics{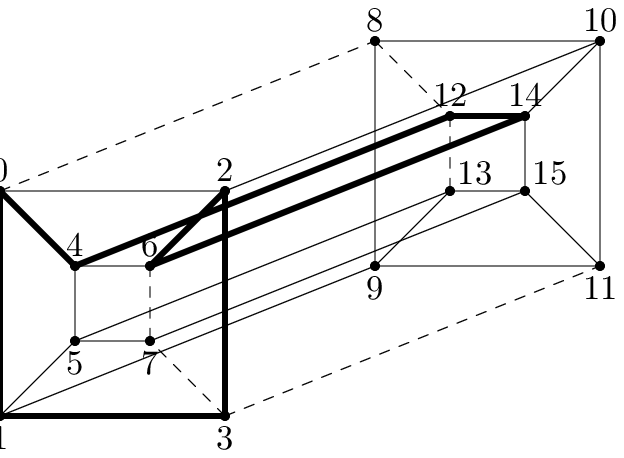}
\caption{The cycle $C_8^3$. The vertices of parity 0 are cut off. If we cut off the vertices with parity 1, then we obtain the isomorphic trap.}
\label{figC8c3trap}
\end{figure}

\begin{figure}[htb]
\centering
\includegraphics{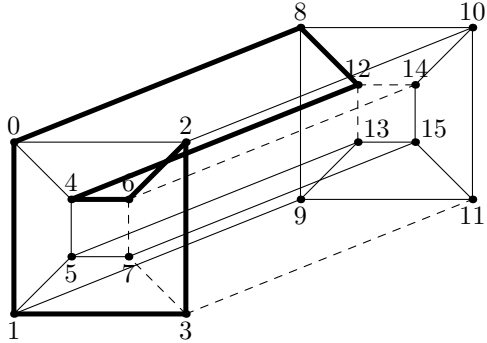}
\caption{The cycle $C_8^4$. The vertices of parity 0 are cut off. If we cut off the vertices with parity 1, then we obtain the isomorphic trap.}
\label{figC8c4trap}
\end{figure}

Case 1. There is a pair of edges of the same dimension separated by one edge.
We may assume that $d_1=d_3=0$ and $d_2=1$. Then there is an edge between $v_0$ and $v_3$, and $(v_0,v_3, v_4,v_5,v_6,v_7)$ is a 3-dim cycle
of length 6. From the discussion in Section~\ref{SecC6} we can deduce 
that there are tree possible sequences of dimensions of $C$:
$(0,1,0,2,3,1,2,3)$, $(0,1,0,2,3,1,3,2)$, or $(0,1,0,2,1,3,2,3)$.
We can assume that the cycles start at 0 and we obtain three cycles:
$C_8^2=(0,1,3,2,6,14,12,8)$, 
 $C_8^3=(0,1,3,2,6,14,12,4)$, and
 $C_8^4=(0,1,3,2,6,4,12,8)$, see Fig.~\ref{figC8c2trap},~\ref{figC8c3trap},~\ref{figC8c4trap}.

\begin{figure}[htb]
\centering
\includegraphics{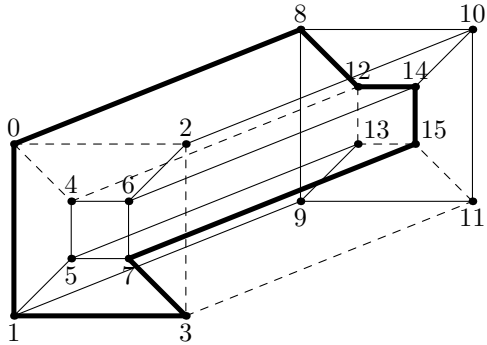}
\caption{The cycle $C_8^5$. The vertices of parity 0 are cut off. If we cut off the vertices with parity 1, then we obtain the isomorphic trap.}
\label{figC8c5trap}
\end{figure}

\begin{figure}[htb]
\centering
\includegraphics{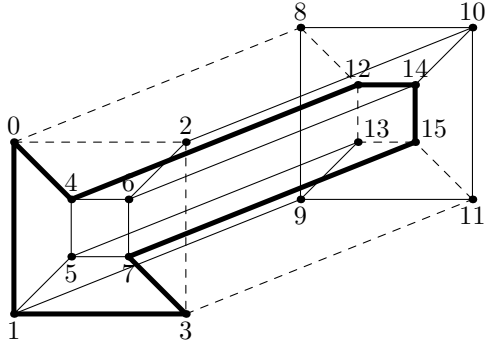}
\caption{The cycle $C_8^6$. The vertices of parity 0 are cut off. If we cut off the vertices with parity 1, then we obtain the trap which is not isomorphic.}
\label{figC8c6trap}
\end{figure}

Case 2. There are no pairs separated by a single edge, but there is a pair
separated by three edges. 
We may assume that the first five dimensions are $0,1,2,3,0$,
and we have  three possible sequences of dimensions in $C$,  namely:
$(0,1,2,3,0,1,2,3)$, $(0,1,2,3,0,1,3,2)$, or $(0,1,2,3,0,2,1,3)$, where
the second and the third sequences are isomorphic.
If we start these cycles at 0, we get two new nonisomorphic cycles of length 8:
 $C_8^5=(0,1,3,7,15,14,12,8)$, and 
$C_8^6=(0,1,3,7,15,14,12,4)$, see Fig.~\ref{figC8c5trap},~\ref{figC8c6trap}.

\begin{figure}[htb]
\centering
\includegraphics{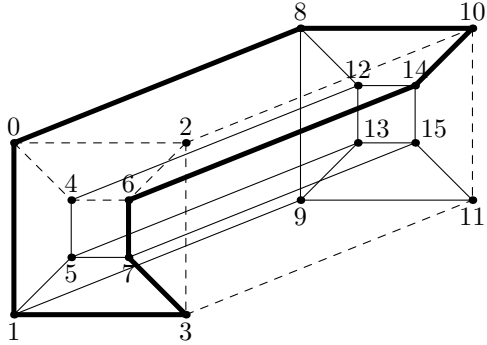}
\caption{The cycle $C_8^7$. The vertices of parity 0 are cut off. If we cut off the vertices with parity 1, then we obtain the trap which is not isomorphic.}
\label{figC8c7trap}
\end{figure}

Case 3. For each $i\in\{0,1,2,3\}$, the pair of edges of dimension $i$
is separated by two (or four) other edges. We have only one possible sequence of dimensions $(0,1,2,0,3,2,1,3)$
and one nonisomorphic cycle
$C_8^7=(0,1,3,7,6,14,10,8)$, see Fig.~\ref{figC8c7trap}. 

Thus, we have seven nonisomorphic cycles of length 8 in $Q_4$:

\begin{itemize}
\item $C_8^1=(0,1,3,2,6,7,5,4)$ which is a Hamiltonian cycle in 3- dimensional subcube $Q_3$ and we have a $Q_3$-DHW trap.
\item $C_8^2=(0,1,3,2,6,14,12,8)$, see Fig~\ref{figC8c2trap}. We need $4n-9$ faulty edges to cut off the vertices 0, 3, 6, and 12 from the vertices outside the cycle.
Note that the vertex 4  is of degree one in $Q_4$. But if we embed
the trap in $Q_n$ with $n\ge 5$, then $4$ has degree at least two.
We get an isomorphic trap,  if we cut off the vertices 1, 2, 14, and 8.
\item $C_8^3=(0,1,3,2,6,14,12,4)$, see Fig~\ref{figC8c3trap}. 
We need $4n-10$ faulty edges to cut off the vertices 0, 3, 6, and 12 from the vertices outside the cycle.
We get an isomorphic trap, if we cut off the vertices 1, 2, 14, and 4.
\item  $C_8^4=(0,1,3,2,6,4,12,8)$, see Fig~\ref{figC8c4trap}. 
We need $4n-10$ faulty edges to cut off the vertices 0, 3, 6, and 12.
Similarly, we need $4n-10$ faulty edges to cut off the vertices 1, 2, 4, and 8.
These two traps are not isomorphic.
\item $C_8^5=(0,1,3,7,15,14,12,8)$, see Fig~\ref{figC8c5trap}. 
We need $4n-8$ faulty edges to cut off the vertices 0, 3, 15, and 12.
We get an isomorphic trap, if we cut off the vertices 1, 7, 14, and 8.
\item $C_8^6=(0,1,3,7,15,14,12,4)$, see Fig~\ref{figC8c6trap}. 
We need $4n-8$ faulty edges to cut off the vertices 0, 3, 15, and 12.
In $Q_4$ we have 8 faulty edges that form a cycle of length 8 and
this is not a minimal trap because the graph is not Hamiltonian even if we change any one of these faulty edges into a healthy edge.
If we embed the trap in $Q_n$ with $n\ge 5$, and add faulty edges in dimensions $i$, $i\ge 5$ to cut off vertices 0, 3, 15 and 12, then the trap is minimal. 

We need $4n-8$ faulty edges to cut off the vertices 1, 7, 14, and 4. In this case the vertices 5 and 6 are 
of degree 1 in $Q_4$. Hence, this trap is not minimal in $Q_4$, but is minimal, for $n\ge5$.
These two traps are not isomorphic.
\item $C_8^7=(0,1,3,7,6,14,10,8)$, see Fig~\ref{figC8c7trap}. 
We need $4n-8$ faulty edges to cut off the vertices 0, 3, 6, and 10. In $Q_4$ the vertex 2 is of degree 0.
Hence, this trap is not minimal in $Q_4$ and $Q_5$, but is minimal, for $n\ge6$.

We need $4n-8$ faulty edges to cut off the vertices 1, 7, 14, and 8.
In $Q_4$ these eight faulty edges form a cycle. Note that in $Q_4$ a set of four of these edges cut off the vertices 9 and 15
from the vertices outside the cycle $(9,11,15,13)$. Similarly, another set of four of these  edges  cut off the vertices 5 and 12
from the vertices outside the cycle $(4,5,13,12)$. Hence, this trap is not minimal in $Q_4$, but is minimal, for $n\ge5$.
These two traps are not isomorphic.
\end{itemize}

\section{Heuristic}\label{H}
In this section we present a heuristic that recognizes some non-Hamiltonian cubes. It is not perfect and does not detect all nonhamiltonian cubes.
The heuristic colors in blue some of the edges in $Q_n$ and removes (adds to the set $F$) some other edges. An edge $e$ is colored in blue, if the heuristic detects that each Hamiltonian cycle in $Q_n$ contains $e$, and an edge $e$ is removed, if it is detected that no Hamiltonian cycle contains $e$. 

The heuristic is based on a few observations.
First observation is about vertices of degree two, i.e. vertices
with only two neighbors.
We shall say that a vertex $v$ is a neighbor of a vertex $u$
only if $u$ and $v$ are neighbors in $Q_n$ and 
the edge $(u,v)\notin F$ is healthy.
Consider a vertex $u$ with only two neighbors $v$ and $w$.
Then the edges $(u,v)$ and $(u,w)$ are in any Hamiltonian cycle
in the cube and we color these edges in blue. Furthermore, if two
blue edges meet in one vertex $x$, then all other edges adjacent to $x$ can be removed, because they do not belong to any Hamiltonian cycle. If three or more blue edges meet in one vertex, then no Hamiltonian cycle is possible.

The second observation is about crossing edges given by partition of $Q_n$ along some dimension $i$.
There are $2^{n-1}$ crossing edges:
half of them are even (their ends in $Q^L$ are of parity  zero)
and the other half are odd (their ends in $Q^L$ are of parity one).
Let
$f_i^e$ denote the number of even faulty crossing edges,
$f_i^o$ denote the number of odd faulty crossing edges,
$b_i^e$ denote the number of blue even crossing edges,
and
$b_i^e$ denote the number of blue odd crossing edges.

\begin{lem}\label{partition}
\begin{enumerate}
\item If $f_i^e=2^{n-2}$ (or $f_i^o=2^{n-2}$), then there is no Hamiltonian cycle
\item If $f_i^e+b_i^o>2^{n-2}$ or $f_i^o+b_i^e>2^{n-2}$, then there is no Hamiltonian cycle
\item If $f_i^e+b_i^o=2^{n-2}$, then we can color in blue all even edges that are  not in $F$ and remove all odd edges that are not blue.
\item If $f_i^o+b_i^e=2^{n-2}$, then we can color in blue all odd edges that are 
not in $F$ and remove all even edges that are not blue.
\item If $f_i^e=2^{n-2}-1$ (or $f_i^o=2^{n-2}-1$), then we can color in blue the only even (or odd, respectively) edge that is not faulty.
\end{enumerate}
\end{lem}

\noindent{\bf Proof:} In the case (1) the subcube $Q^L$ is a trap disconnected halfway.

Case (2). Suppose for a contradiction that there is a Hamiltonian cycle $C$. By Lemma~\ref{cycle2}, the set of crossing edges that are in $C$ must be balanced. Hence, it must contain at least $b^o$ even edges. This is not possible, because there are  only $2^{n-2}-f_i^e<b^o$ healthy even edges.

Case (3). 
Suppose that $C$ be an arbitrary Hamiltonian cycle, and let $B_i^e$ denote the set of even crossing edges that belong to $C$. On the one hand, we have
$|B_i^e|\le 2^{n-2}- f_i^e$, because $B_i^e\cap F=\emptyset$.
On the other hand, $|B_i^e|\ge b_i^o$, because the set of crossing edges in $C$
is balanced. Thus, $|B_i^e|=2^{n-2}-f_i^e$ and all healthy even crossing edges must belong to $C$. 
Furthermore, let $B_i^o$ denote the set of odd crossing edges that belong to $C$.
Because the set of crossing edges in $C$ is balanced, we have
$|B_i^o|=|B_i^e|=2^{n-2}-f_i^e=b_i^o$. Hence, all edges in $B_i^o$ are already
colored in blue. Thus, we have proved that all healthy
even crossing edges belong to every Hamiltonian cycle (and can be colored in blue) and that only odd edges which are colored in
blue can be in any Hamiltonian cycle (so the odd edges that are not blue
can be removed).

The proof of case (4) is symmetric to the proof of (3).

Case (5) Every Hamiltonian cycle must have at least one even and one odd edge in each dimension.
\hfill$\Box$

\bigbreak
The third observation is about blue cycles. If we get a blue cycle which is shorter than $2^n$, then there are no Hamiltonian cycle in $Q_n$.
Similarly, we can remove the edge $e$ if after coloring $e$ in blue, we get a blue cycle shorter than $2^n$.
\\

\renewcommand{\algorithmicrequire}{\textbf{Input:}}
\renewcommand{\algorithmicensure}{\textbf{Output:}}
\begin{algorithm}[h!t]
\begin{algorithmic}[1]
\Require a cube $Q_n$ and a set of faulty edges $F$.
\Ensure the answer ``no HC'', if heuristic detects that there are no HC in $Q_n$
or the answer ``?'' (``do not know''), if heuristic does not find a proof that $Q_n$ is not Hamiltonian
\Repeat
  
\For{each vertex $v$ of degree 2} 
\State \parbox[t]{\dimexpr\linewidth-\algorithmicindent-\algorithmicindent}{color in blue the two healthy edges adjacent to $v$,\strut}
\EndFor

\For{each vertex $v$ with adjacent two blue edges} 
\State{remove all other edges adjacent  to $v$} 
\EndFor

\For{each non blue edge $e$} 
\If{there is a nonhamiltonian cycle containing edge $e$ and blue edges,}
\State{remove edge $e$}
\EndIf 
\EndFor

\If{there is a vertex with three or more adjacent blue edges, }
\State{\Return "no HC"}
\EndIf

\If{there is a partition satisfying the condition (1) or (2) in Lemma~\ref{partition},}
\State{\Return "no HC"}
\EndIf

\If{there is a partition satisfying the condition (3), (4) or (5) in Lemma~\ref{partition},} 
\State{color or remove edges according to the lemma.}
\EndIf

\Until{there are no edges to color or remove}

\If{there is a trap of the kinds  described in the previous sections,} 
\State{\Return "no HC"}
\EndIf

\State{\Return "?"}
\end{algorithmic}

\label{alg:test}
\end{algorithm}




Note that the heuristic detects traps of the kinds  described in the previous sections.
The heuristic is not perfect, as we shall see later there are nonhamiltonian 4-cubes that are not detected by 
the heuristic.

\section{Hamiltonian cycles in $Q_4$}\label{Q4}
In  $Q_4$ we can build the following traps of the kinds mentioned in the previous sections, see Table~\ref{tab2} and~\ref{tab3}:
\begin{itemize}
\item with 3 faulty edges, the $Q_1$-DHW trap (vertex of degree one),
\item with 4 faults, the $Q_2$-DHW, or $Q_3$-DHW trap,
\item with 5 faults, the $C_6^1$-DHW trap,
\item with 6 faults, the $CL$ trap, $C_8^3$-DHW, or $C_8^4$-DHW. 
We can also build the $C_6^2$-DHW but this trap is not minimal, because it contains a $Q_1$-DHW trap, see Section~\ref{SecC6},
\item with 7 faults the $C_8^2$-DHW but this trap is not minimal, it contains $Q_1$-DHW trap,
\item with 8 faults, the $C_8^5$-DHW, $C_8^6$-DHW, or $C_8^7$-DHW trap. In the second or the third case we obtain traps which are not minimal, see Section~\ref{C8}.
\end{itemize}

From the results of ~\cite{3,4,Sz} it follows  that  $Q_4$ is Hamiltonian, if (see Table~\ref{tab1}):
\begin{itemize}
\item it does not contain more than $2$ faulty edges,
\item it has  $3$ faulty edges and all its vertices are of degree at least 2, or
\item it has $4$ faulty edges and contains neither $Q_1$-DHW, $Q_2$-DHW, nor $Q_3$-DHW trap.
\end{itemize}

We performed computer experiments and found out that:
\begin{itemize}
\item with  5 faulty edges $Q_4$ is not Hamiltonian
only if it contains a $Q_1$-DHW, $Q_2$-DHW, $Q_3$-DHW, or $C_6^1$-DHW trap,
\item with  6 faulty edges $Q_4$ is not Hamiltonian
only if it contains a  $Q_1$-DHW, $Q_2$-DHW, $Q_3$-DHW, $C_6^1$-DHW,  $C_8^3$-DHW, $C_8^4$-DHW, or $CL$ trap.
\end{itemize}

There are 25 nonisomorphic $Q_4$ cubes with 7  faulty edges that are not Hamiltonian and do not contain any traps of the kinds described above.
All these cubes can be recognized by the heuristic described in the previous section.
There are nonhamiltonian 4-cubes that are not detected by 
the heuristic. More precisely:
\begin{itemize}
\item there are five such minimal cubes with 8 faulty edges, see Appendix, 
\item there are eight such minimal cubes with 9 faulty edges, see Appendix, 
\item there are  four such minimal cubes with 10 faulty edges, see Appendix,
\item there are no such minimal cubes with eleven or more faulty edges.
\end{itemize}

\section{Hamiltonian cycles in $Q_5$}\label{Q5}
In $Q_5$ we can build the following traps of the kinds mentioned above:
\begin{itemize}
\item with 4 faulty edges, the $Q_1$-DHW trap (vertex of degree one),
\item with 6 faults, the $Q_2$-DHW trap,
\item with 8 faults, the  $Q_3$-DHW, $Q_4$-DHW, or $C_6^1$-DHW trap,
\item with 9 faults, the $CL$, or $C_6^2$-DHW trap,
\item with 10 faults, the $C_8^3$-DHW, or $C_8^4$-DHW trap, 
\item with 11 faults the $C_8^2$-DHW trap,
\item with 12 faults, the $C_8^5$-DHW, $C_8^6$-DHW, $C_8^7$-DHW trap.
\end{itemize}

From~\cite{3,4,Sz,Liu} it follows  that  $Q_5$ is Hamiltonian, if (see Table~\ref{tab1}):
\begin{itemize}
\item it does not contain more than $3$ faulty edges,
\item it has  $4$ or $5$ faulty edges and all of its vertices are of degree at least 2,
\item it has $6$ or $7$ faulty edges and contains neither $Q_1$-DHW, nor $Q_2$-DHW    trap.
\end{itemize}

We performed computer experiments and found that:

\begin{itemize}
\item with  8 faulty edges $Q_5$ is not Hamiltonian
only if it contains a $Q_1$-DHW, $Q_2$-DHW, $Q_3$-DHW, $Q_4$-DHW, or $C_6^1$-DHW trap,
\item with  9 faulty edges $Q_4$ is not Hamiltonian
only if it contains  $Q_1$-DHW, $Q_2$-DHW, $Q_3$-DHW, $Q_4$-DHW, $C_6^1$-DHW, $C_6^2$-DHW, or a $CL$ trap.
\end{itemize}
We are not able to give a  full description of the $Q_5$ cubes with ten faulty edges that
are not Hamiltonian. As we mentioned above, with ten faulty edges we can build a trap disconnected halfway with a cycle of length 8.
Moreover with ten faulty  edges we can build in $Q_5$ three other types of traps
disconnected halfway.

\begin{figure}[htb]
\centering
\includegraphics{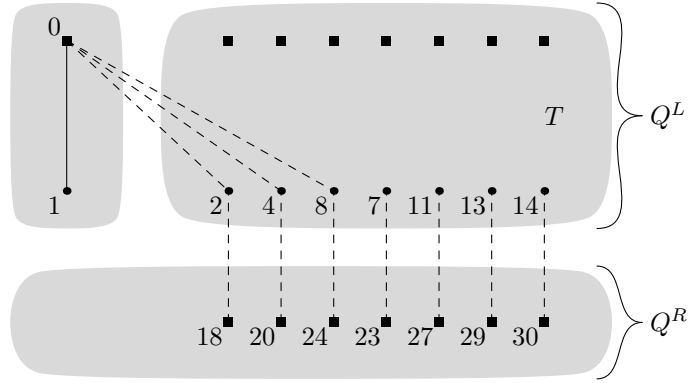}
\caption{The set $T$ disconnected halfway. The vertices of parity zero are depicted by squares,
the vertices of parity one are depicted by cycles. The vertices in $T$ of parity one belong to $T^1$.}
\label{figTtrap}
\end{figure}

Consider partition of $Q_5$ along dimension 4 into two 4-dimensional
subcubes
$Q^L=\{x\mid 0\le x\le 15\}$ and
$Q^R=\{x\mid 16\le x\le 31\}$.
In $Q^L$ we have 1-dimensional subcube $Q_1=\{0,1\}$
and with three faulty edges:
$(0,2)$,
$(0,4)$,
and $(0,8)$ we can cut off the vertex 0 (the only vertex in $Q_1$
with parity 0) from the set $T=Q^L-Q_1=\{x\in Q_5\mid 2\le x\le15\}$, see Fig.~\ref{figTtrap}.

Note that the same three faulty edges cut off the set of vertices in $T$ of parity one, denoted by $T^1$, from the vertices in $Q_1$.
Hence, the three edges cause that both $Q_1$ and $T$ are disconnected halfway in $Q^L$. 
Moreover, with seven faulty edges
$(2,18)$, $(4,20)$, $(7,23)$, $(8,24)$, $(11,27)$, $(13,29)$, $(14,30)$ we can cut off $T^1$ from $Q^R$.
Hence, with ten faulty edges
$(0,2)$,
$(0,4)$,
$(0,8)$,
$(2,18)$, $(4,20)$, $(7,23)$, $(8,24)$, $(11,27)$, $(13,29)$, $(14,30)$ we can cut off $T^1$ from the set $Q_5-T$. Thus,
$T$ is disconnected halfway and $Q_5$ is not Hamiltonian. Note that the vertices of $T$ can be arranged in a cycle of length 14.

\begin{figure}[htb]
\centering
\includegraphics{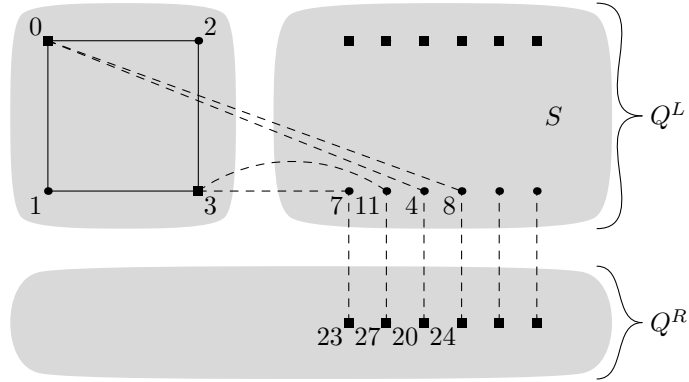}
\caption{The set $S$ disconnected halfway. The vertices of parity zero are depicted by squares,
the vertices of parity one are depicted by cycles. The vertices in $S$ of parity one belong to $S^1$.}
\label{figStrap}
\end{figure}

Similarly, consider the 2-dimensional subcube $Q_2=\{0,1,2,3\}$.
With four faulty edges:
$(0,4)$,
$(0,8)$,
$(3,7)$,
$(3,11)$
we can cut off the vertices 0 and 3 (the vertices in $Q_2$
with parity 0) from the set $S=Q^L-Q_2=\{x\in Q_5\mid 4\le x\le 15\}$, see Fig.~\ref{figStrap}.
The same four faulty edges cut off the set of vertices in $S$ of parity one, denoted by $S^1$, from the vertices in $Q_2$.
Hence, the four edges cause that both $Q_2$ and $S$ are disconnected halfway in $Q^L$. 
Moreover, with six faulty edges
$(4,20)$, $(7,23)$, $(8,24)$, $(11,27)$, $(13,29)$, $(14,30)$
 we can cut off $S^1$ from $Q^R$.
Hence, with ten faulty edges
$(0,4)$,
$(0,8)$,
$(3,7)$,
$(3,11)$,
$(4,20)$, $(7,23)$, $(8,24)$, $(11,27)$, $(13,29)$, $(14,30)$
we can cut off $S^1$ from the set $Q_5-S$. Thus,
$S$ is disconnected halfway in $Q_5$. The vertices of $S$ can be arranged in a cycle of length 12.

\begin{figure}[htb]
\centering
\includegraphics{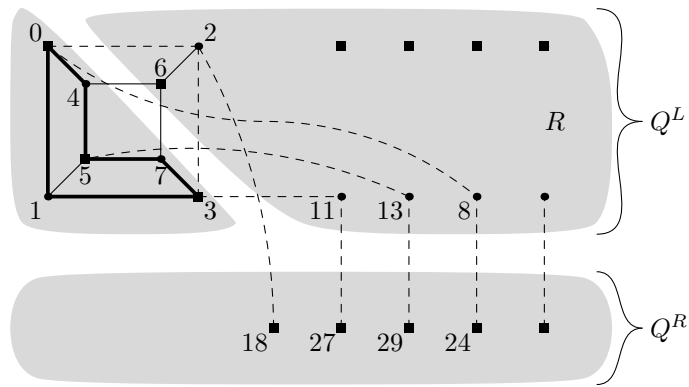}
\caption{The set $R$ disconnected halfway. The vertices of parity zero are depicted by squares,
the vertices of parity one are depicted by cycles. The vertices in $R$ of parity one belong to $R^1$.}

\label{figRtrap}
\end{figure}

In order to obtain the third trap with ten faulty edges let us consider
the cycle $C_6^1=\{0,1,3,7,5,4\}$.
With five faulty edges
$(0,2)$,
$(2,3)$,
$(5,13)$,
$(3,11)$, $(0,8)$
we can cut off the vertices $\{0,3,5\}$ (the vertices in $C_6^1$ of parity zero) from the set $R=Q^L-C_6^1=\{x\in Q_5\mid 7<x<16\}\cup \{2,6\}$, see Fig.~\ref{figRtrap}.
The same five faulty edges cut off the set of vertices in $R$ of parity one, denoted by $R^1$, from the vertices in $C_6^1$.
With seven more faulty edges $(3,11)$, $(0,8)$, 
$(2,18)$, 
$(8,24)$, $(11,27)$, $(13,29)$, $(14,30)$
we can cut off $R^1$ from $Q^R$.
Hence, with ten faulty edges
$(0,2)$,
$(2,3)$,
$(5,13)$,
$(3,11)$, $(0,8)$, 
$(2,18)$, 
$(8,24)$, $(11,27)$, $(13,29)$, $(14,30)$
we can cut off $R^1$ from the set $Q_5-R$. Thus,
$R$ is disconnected halfway in $Q_5$. The vertices of $R$ can be arranged in a cycle of length 10.

\section{Remarks and open problems}\label{open}
We applied the ideas used in the heuristic, described in Section 6, to speed up the simple backtracking algorithm which looks for Hamiltonian cycle in $Q_n$. We used the new algorithm to detect all nonhamiltonian 4- dimensional cubes and all nonhamiltonian 5- dimensional cubes with 8 or 9 faulty edges. It worked much quicker than the simple backtracking algorithm.

We conjecture that each cube $Q_n$ with $n\ge 6$ and with  $3n-6$ faulty edges  is not Hamiltonian
only if it contains a $Q_1$-DHW, $Q_2$-DHW, $C_6$-DHW, or $CL$ trap.


\section{Appendix}

There are $Q_4$ cubes with 8, 9 or 10 faulty edges that are not Hamiltonian, do not contain any local traps of the kinds described in this paper, and cannot be detected by the heuristic. Below we present all such cubes.

\begin{enumerate}
\item There are five minimal cubes with 8 faulty edges. Set $F$ contains edges $(0,1)$, $(7,15)$ and egdes: 
\begin{enumerate}
\item $(10,11)$, $(1,3)$, $(12,14)$, $(2,6)$, $(8,12)$, $(0,8)$; 
\item $(10,11)$, $(1,3)$, $(4,6)$, $(2,6)$, $(8,12)$, $(2,10)$; 
\item $(2,3)$, $(0,2)$, $(13,15)$, $(3,7)$, $(8,12)$, $(5,13)$; 
\item $(2,3)$, $(4,5)$, $(8,9)$, $(0,2)$, $(12,14)$, $(11,15)$; 
\item $(2,3)$, $(4,5)$, $(8,9)$, $(0,2)$, $(12,14)$, $(1,9)$.
\end{enumerate}

\item There are eight minimal cubes with 9 faulty edges. Set $F$ contains edges $(0,1)$, $(2,3)$, $(7,15)$ and egdes:
\begin{enumerate}
\item $(12,13)$, $(9,11)$, $(12,14)$, $(1,5)$, $(10,14)$, $(0,8)$; 
\item $(12,13)$, $(0,2)$, $(9,11)$, $(8,12)$, $(10,14)$, $(5,13)$; 
\item $(12,13)$, $(4,6)$, $(8,10)$, $(8,12)$, $(10,14)$, $(5,13)$; 
\item $(4,5)$, $(9,11)$, $(0,4)$, $(1,5)$, $(10,14)$, $(6,14)$; 
\item $(4,5)$, $(0,2)$, $(13,15)$, $(8,12)$, $(5,13)$, $(6,14)$; 
\item $(8,9)$( $(12,13)$, $(4,6)$, $(3,7)$, $(10,14)$, $(0,8)$; 
\item $(4,5)$, $(6,7)$, $(8,9)$, $(12,14)$, $(11,15)$, $(6,14)$; 
\item $(4,5)$, $(10,11)$, $(0,2)$, $(1,3)$, $(12,14)$, $(9,13)$. 

\end{enumerate}

\item There are four minimal cubes with 10 faulty edges. Set $F$ contains edges $(0,1)$, $(2,3)$, $(4,5)$, $(7,15)$ and egdes:
\begin{enumerate}
\item $(0,2)$, $(1,3)$, $(8,10)$, $(9,13)$, $(11,15)$, $(6,14)$; 
\item $(1,3)$, $(8,10)$, $(9,13)$, $(11,15)$, $(4,12)$, $(6,14)$; 
\item $(4,6)$, $(8,10)$, $(12,14)$, $(10,14)$, $(11,15)$, $(3,11)$; 
\item $(10,11)$, $(12,14)$, $(0,4)$, $(2,6)$, $(9,13)$, $(3,11)$. 
\end{enumerate}
\end{enumerate}

\end{document}